%% file: main.tex
\documentclass[11pt]{article}

\usepackage[T1]{fontenc}
\usepackage[utf8]{inputenc}
\usepackage{lmodern}
\usepackage[margin=1in]{geometry}
\usepackage{microtype}
\usepackage{amsmath,amssymb}
\usepackage{booktabs}
\usepackage{tabularx}
\usepackage{array}
\usepackage{xcolor}
\usepackage{graphicx}
\usepackage{tikz}
\usetikzlibrary{arrows.meta,positioning,fit,backgrounds}
\usepackage[section]{placeins}
\usepackage{flafter}
\usepackage{float}
\usepackage{enumitem}
\usepackage{titlesec}
\usepackage{url}
\usepackage[hidelinks]{hyperref}
\usepackage[nameinlink,noabbrev]{cleveref}
\usepackage{claims}

\input{generated/results}

\definecolor{approvalblue}{HTML}{1F5A94}
\definecolor{attackred}{HTML}{A12A2A}
\definecolor{safegreen}{HTML}{257A4B}
\definecolor{softgray}{HTML}{F2F4F7}
\newcommand{\A}{\ensuremath{A}}
\newcommand{\B}{\ensuremath{B}}

\newcommand{\code}[1]{\texttt{#1}}

\newcolumntype{Y}{>{\raggedright\arraybackslash}X}
\newcolumntype{P}[1]{>{\raggedright\arraybackslash}p{#1}}
\titleformat{\section}{\normalfont\Large\bfseries}{\thesection}{0.75em}{}
\titleformat{\subsection}{\normalfont\large\bfseries}{\thesubsection}{0.75em}{}

\title{Loopjacking: Hijacking Human-in-the-Loop Approval}
\author{Adithyan Arun Kumar\\
Independent Security Researcher\\
\href{mailto:adioffsec@gmail.com}{\texttt{adioffsec@gmail.com}}}
\date{September 2026}
\hypersetup{
  pdftitle={Loopjacking: Hijacking Human-in-the-Loop Approval},
  pdfauthor={Adithyan Arun Kumar},
  pdfsubject={Human-in-the-loop approval binding in agent workflows},
  pdfkeywords={agent security, human-in-the-loop, approval, authorization, Loopjacking}
}

\begin{document}
\maketitle

\input{sections/introduction}
\input{sections/model}
\input{sections/cases}
\input{sections/evaluation}
\input{sections/related-work}
\input{sections/defenses}
\input{sections/limitations}
\input{sections/conclusion}
\input{sections/ai-use}

\bibliographystyle{plain}
\begingroup
\small
\raggedright
\bibliography{references}
\endgroup

\end{document}

%% file: generated/results.tex
\newcommand{\AgnoAttackObserved}{5}
\newcommand{\AgnoAttackTotal}{5}
\newcommand{\AgnoDirectBDenied}{3}
\newcommand{\AgnoDirectBTotal}{3}
\newcommand{\AgnoCompletedTrials}{23}

\newcommand{\OpenClawAffectedObserved}{3}
\newcommand{\OpenClawAffectedTotal}{3}
\newcommand{\OpenClawAffectedDirectBDenied}{3}
\newcommand{\OpenClawFixedObserved}{3}
\newcommand{\OpenClawFixedTotal}{3}
\newcommand{\OpenClawFixedDirectBDenied}{3}
\newcommand{\OpenAIEarlierHonestPass}{3}
\newcommand{\OpenAIEarlierHonestTotal}{3}
\newcommand{\OpenAIEarlierMutationRejected}{3}
\newcommand{\OpenAIEarlierMutationTotal}{3}
\newcommand{\OpenAICurrentHonestPass}{3}
\newcommand{\OpenAICurrentHonestTotal}{3}
\newcommand{\OpenAICurrentMutationRejected}{3}
\newcommand{\OpenAICurrentMutationTotal}{3}

%% file: sections/introduction.tex
\begin{abstract}
Human approval is often treated as the last security boundary before an agent
executes a consequential operation. That boundary is only meaningful if the
operation presented for review is the operation later authorized or released.
We call failures of this binding \emph{Loopjacking}: a human approves what they
understand as operation \A, while the implementation uses that decision for a
materially different operation \B. We distinguish two variants. In a
representation-based attack, \B\ is already encoded but omitted or misrepresented
at approval time; in a post-approval state-substitution attack, the human sees the
correct \A\ and mutable workflow state later replaces it with \B.

We evaluate a purposive set of released agent products. We reproduce
post-approval substitution in seven tested Agno AgentOS releases ending at 3.0.9
and in 12 tested versions of a conditional in-memory LangGraph Agent Server
composition ending at 0.14.0. We reproduce representation mismatch in OpenClaw
2026.2.23 and its rejection in 2026.2.24. OpenAI Agents SDK 0.22.0 and 0.22.2 provide a
negative control: serialized continuation preserves exact per-call binding and
rejects mutated \B. These results do not estimate ecosystem prevalence. They show
that complete canonical approval rendering and exact use-time comparison, or
preventing unauthorized pending-state mutation, block the tested attacks while
preserving legitimate execution. We separate this contribution from established
work on misleading dialogs, session smuggling, action binding, and authorization
continuity.
\end{abstract}

\section{Introduction}
\label{sec:introduction}

A human-in-the-loop prompt can look like a decisive security control: an agent
proposes an operation, a person inspects it, and execution proceeds only after an
explicit decision. In practice, the reviewed operation may pass through several
representations. A product can render one string, persist a structured tool call,
accept a continuation message, reconstruct a current action, and finally dispatch
different arguments to a sink. Approval is sound only when these stages agree on
the material effect.

That agreement is not guaranteed by the presence of an approval button. Consider
a low-privilege user who can request work but cannot authorize a protected
transfer. An administrator reviews a transfer of 20 units to an approved vendor
and approves it. If the user can then replace the pending arguments with a
transfer of 2,000 units to an attacker-controlled destination, and the product
executes the replacement under the administrator's still-effective decision, the
human participated but the control failed. A second form needs no later mutation:
the complete request already encodes the harmful command, while the approval view
shows only an incomplete shell fragment.

We use \emph{Loopjacking} for this product-owned approval-to-effect mismatch. A
qualifying trace requires a genuine human decision understood as \A, materially
different \B, reachable attacker influence, product-owned consumption of the
decision for \B, a consequential sink, and evidence that the attacker lacked an
equivalent direct path to \B. Representation mismatch and post-approval state
substitution are distinct variants of the same failed binding.\claim{C101} The
definition deliberately excludes ordinary persuasion where the human knowingly
approves visible \B, generic mutable state, confirmation forgery without a human
decision, and prompt injection that never reuses an approval.

This distinction matters for both testing and repair. Looking only for mutable
tasks misses commands that were misrepresented before approval. Looking only at
the dialog misses a correct review whose underlying operation changes afterward.
Conversely, treating every continuation or policy bypass as Loopjacking would make
the term too broad to identify the failed security boundary.

This paper makes four contributions:

\begin{enumerate}[leftmargin=*,itemsep=2pt]
  \item It gives a testable model of the human-decision-to-authorization link and
        separates representation-based from post-approval state-substitution
        traces.
  \item It reports controlled native evaluations of three released product paths:
        an Agno AgentOS configuration, a conditional LangGraph Agent Server
        composition, and OpenClaw's shell-wrapper approval path.
  \item It includes OpenAI Agents SDK ordinary per-call approval as a negative
        control, showing that resumability and serialization alone do not imply
        stale approval reuse.
  \item It validates two complementary defenses on the tested traces: complete
        canonical presentation with exact use-time binding, and authorization
        policies that prevent an untrusted actor from modifying pending approved
        state.
\end{enumerate}

The study is comparative, not representative. Its population was selected to test
mechanically different approval paths and a falsifying control. We report exact
versions, configurations, trial denominators, and boundaries; we do not infer a
market-wide rate, a universal affected range, or one severity score. The evidence
cutoff is September 10, 2026.

%% file: sections/model.tex
\section{Operations, Approvals, and Attacker Model}
\label{sec:model}

\subsection{Approval-mediated operations}

Approval-mediated execution has two security-relevant moments. At approval time,
the product presents a human-visible view of the proposed operation. The complete
operation includes the action, arguments, target resource, principal and task
scope, and any execution context that can materially change the effect. The human
understands the displayed operation as \A\ and records decision \(D\). At use time,
a later component applies \(D\) to the current operation, which may no longer be
\A.

\noindent\textbf{Approval-binding invariant.}
A decision may authorize an operation only when the complete operation evaluated
at use time is materially equivalent to what the human reviewed and approved, and
the decision remains valid for the current principal, task, and scope. The
enforcement point must reconstruct the operation after all attacker-influenced
transformations. If the operation differs materially, the product must reject it
or obtain a new approval. A human may intentionally approve an entire mutable task,
but only when that breadth is clearly presented and the resulting decision still
constrains every later effect.

\begin{figure}[t]
  \centering
  \resizebox{\textwidth}{!}{\input{figures/approval-model}}
  \caption{The approval-binding model. Loopjacking occurs when the product applies
  a decision over human-visible \A\ to release materially different \B. The safe
  branch reconstructs the complete current operation and either executes unchanged
  \A\ or rejects or reauthorizes \B.}
  \label{fig:approval-model}
\end{figure}
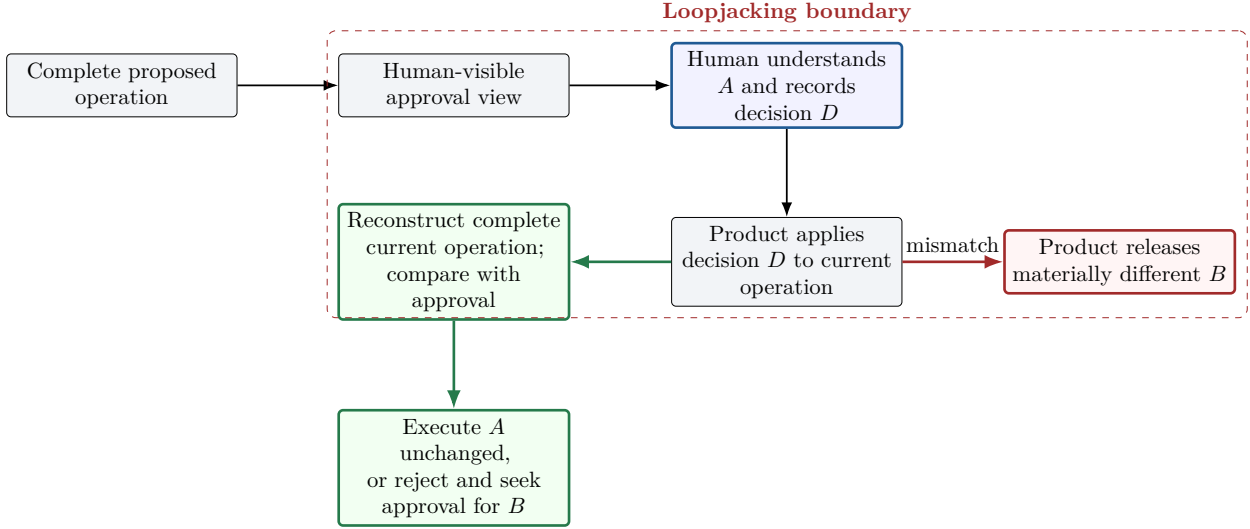
\FloatBarrier

\subsection{Definition and variants}

\textbf{Loopjacking} is an implementation-level failure in which a human approves
the operation or representation they understand as \A, but product-owned logic
uses that decision to authorize or release materially different \B.\claim{C101}
The attacker hijacks the link between a human decision and the operation that
inherits its authority, not necessarily the surrounding conversation or the
human's account.

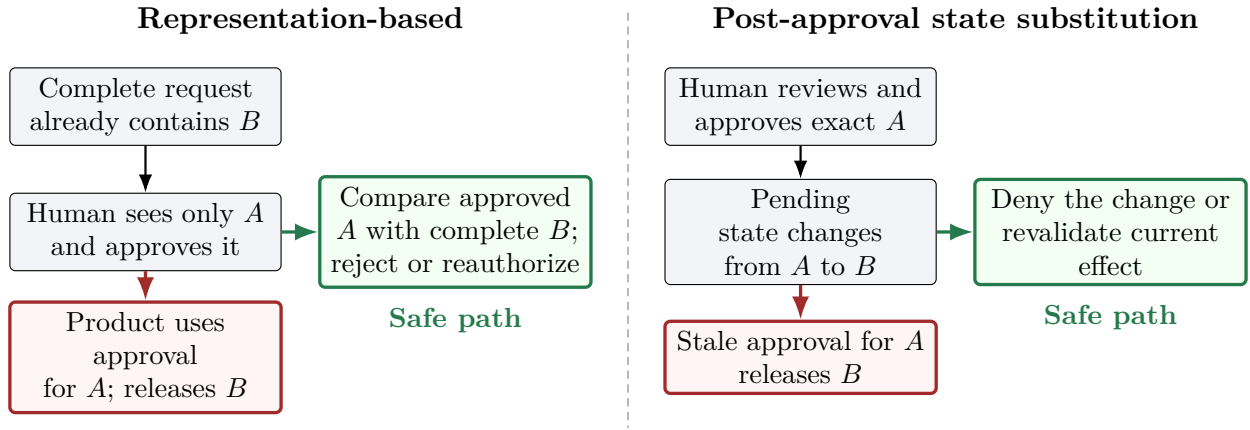
\begin{figure}[H]
  \centering
  \resizebox{\textwidth}{!}{\input{figures/variants}}
  \caption{The two primary variants and their safe branches. In the
  representation variant, \B\ exists before review but the view shows \A. In the
  state-substitution variant, correct \A\ becomes \B\ after review. Complete
  rendering and use-time binding address both; preventing unauthorized mutation
  is an additional control for the state variant.}
  \label{fig:variants}
\end{figure}

The definition yields two primary variants, shown in
\cref{fig:variants}:

\noindent\textbf{Representation-based Loopjacking}\par
\noindent The complete request or execution context contains \B\ before the
decision, but the product displays, canonicalizes, or checks a materially
incomplete \A. The approval is accurate only for the incomplete representation.

\medskip
\noindent\textbf{Post-approval state substitution}\par
\noindent The human sees and approves exact \A. Before the approval is consumed,
an actor changes pending task, thread, session, or continuation state to \B.
Execution evaluates current \B\ while retaining the decision made for \A.

Replay, scope drift, and inconsistent use-time interpretation can fit the same
definition when a real decision for \A\ is what authorizes \B. They are not
automatically Loopjacking merely because an approval record is stale or a workflow
is mutable.

\subsection{Necessary conditions}

We admit a trace only when all six conditions below hold. Together they prevent
the label from collapsing into generic agent compromise.

\begin{enumerate}[leftmargin=*,itemsep=2pt]
  \item \textbf{Human approval.} A person or distinct approval role makes a
        product-recorded decision over a view understood as \A.
  \item \textbf{Material mismatch.} \B\ changes an authorization-relevant effect,
        not merely formatting or an immaterial field.
  \item \textbf{Reachable influence.} An attacker can induce the mismatch through
        a supported or realistically reachable product path.
  \item \textbf{Product-owned consumption.} Product code presents, stores,
        transforms, or consumes the decision in a way that releases \B.
  \item \textbf{Consequential sink.} The exact operation reaching the sink is
        recorded; an intermediate state or model statement is insufficient.
  \item \textbf{No equivalent direct authority.} The attacker cannot obtain the
        same material effect without hijacking the approval.
\end{enumerate}

This operational definition and its two-variant grouping are a research taxonomy,
not a claim that approval integrity or action binding is new. Existing work already
requires trusted presentation and binding between consent and execution
\cite{Weng2026ConsentIntegrity,YuEtAl2026SUDP,Microsoft2026ActionBoundApproval}.

\subsection{Attacker and trust boundaries}

The attacker may be a low-privilege task initiator, a repository writer, a remote
agent, or another actor with a narrow state-update capability. The attacker need
not control the model and need not race the approver. What matters is asymmetric
authority: the attacker can influence the request or pending state but cannot
authorize \B; the approver can authorize consequential execution but does not
intend \B.

We exclude an omnipotent database writer who can forge every trusted approval
field, because such a principal already controls the authorization substrate. We
also exclude cases where \B\ executes before any decision, approval prompts are
skipped entirely, or the only event is ordinary user input or validation. Prompt
injection, memory poisoning, or session smuggling can supply \B, but a Loopjacking
trace begins only when the product binds a decision for \A\ to \B.

Our experiments evaluate system binding, not human susceptibility. The approval
role proceeds only after the test has asserted and recorded the exact product view
of \A. This creates a deterministic approval decision while deliberately avoiding
claims about interface comprehension, deception rates, or user behavior.

\subsection{A2A as a conditional carrier}
\label{sec:a2a-carrier}

Before section 7.6.4, A2A made three coordination choices relevant to a deferred
approval path. \texttt{TASK\_STATE\_AUTH\_REQUIRED} was an interrupted, nonterminal
state; agents were advised to accept messages directed to the same Task while
authorization remained pending; and an agent that received a credential out of
band could continue processing without a client follow-up. The surrounding text
used human approval before a destructive action as one example, but did not define
the approver, a canonical operation, an action-scoped approval object, or the
application sink \cite{A2APre2081}.

\begin{figure}[!htbp]
  \centering
  \resizebox{\textwidth}{!}{\input{figures/a2a-carrier}}
  \caption{A2A as a conditional carrier. Blue boxes are protocol coordination;
  gray boxes are implementation- or issuer-owned semantics. Only the red branch
  is Loopjacking: implementation code consumes decision \(D_A\) for \B. The green
  branch uses the same A2A mechanics but compares the current operation with the
  decision and rejects or reauthorizes the mismatch.}
  \label{fig:a2a-carrier}
\end{figure}
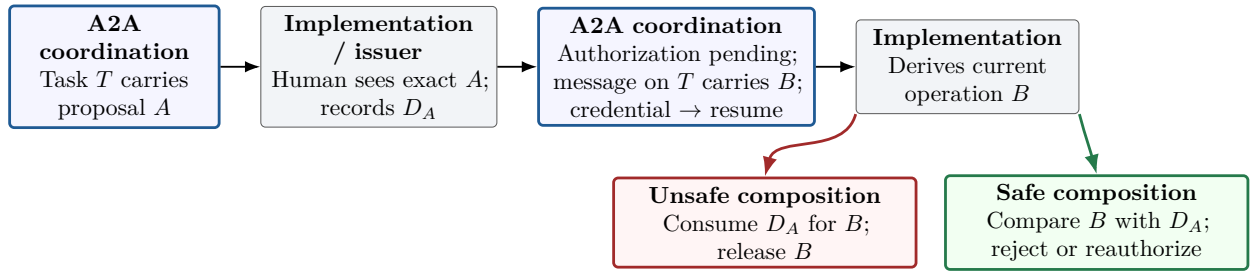

These mechanics do not by themselves create a vulnerability. Consider instead the
conditional composition in \cref{fig:a2a-carrier}. A Task writer who cannot approve
or directly execute \B\ proposes \A. The product presents exact \A\ to a separate
approver and records decision \(D_A\). Before that decision is consumed, the writer
sends \B\ on the same nonterminal Task. If implementation-owned logic selects \B{}
from current Task state after credential receipt and consumes \(D_A\) without a
use-time scope comparison, the product releases \B{} under approval for \A. That
composition is post-approval Loopjacking. A2A supplies the Task, message,
interruption, and resume coordination. The implementation or credential issuer
owns the approval view and decision scope; the implementation owns operation
selection, use-time comparison, and the consequential sink.\claim{C108}

Issue 2080 framed the missing responsibility assignment as implementer ambiguity,
not a request to expand core protocol scope \cite{A2AIssue2080}. Merged pull
request 2081 added section 7.6.4: the interrupted state is a coordination signal,
not an authorization grant, and the implementation, issuer, or extension must
define the authorized operation and check later use \cite{A2APR2081}. At our
September 10, 2026 cutoff, v1.0.1 remained the latest tagged release
\cite{A2ARelease101}. This history supports a specification-clarity lesson, not an
intrinsic A2A or common-SDK vulnerability. The native cases that follow establish
their approval semantics and unsafe consumption independently.\claim{C106}
\FloatBarrier

%% file: figures/approval-model.tex
\begin{tikzpicture}[
  node distance=14mm and 16mm,
  >=Latex,
  every node/.style={font=\small},
  box/.style={draw, rounded corners=2pt, align=center, minimum height=10mm, text width=34mm, fill=softgray},
  decision/.style={box, draw=approvalblue, very thick, fill=blue!5},
  effect/.style={box, draw=attackred, very thick, fill=red!4},
  good/.style={box, draw=safegreen, very thick, fill=green!5},
  flow/.style={->, thick},
  bad/.style={->, very thick, draw=attackred},
  safe/.style={->, very thick, draw=safegreen}
]
  \node[box] (state) {Complete proposed\\operation};
  \node[box, right=of state] (present) {Human-visible\\approval view};
  \node[decision, right=of present] (human) {Human understands\\\(\A\) and records\\decision \(D\)};
  \node[box, below=of human] (consume) {Product applies\\decision \(D\) to current\\operation};
  \node[effect, right=of consume] (effect) {Product releases\\materially different \(\B\)};

  \draw[flow] (state) -- (present);
  \draw[flow] (present) -- (human);
  \draw[flow] (human) -- (consume);
  \draw[bad] (consume) -- node[above]{mismatch} (effect);

  \node[good, left=of consume] (verify) {Reconstruct complete\\current operation;\\compare with approval};
  \node[good, below=of verify] (allow) {Execute \(\A\) unchanged,\\or reject and seek\\approval for \(\B\)};
  \draw[safe] (consume) -- (verify);
  \draw[safe] (verify) -- (allow);

  \node[draw=attackred, dashed, rounded corners, fit=(present)(human)(consume)(effect), inner sep=5pt,
        label={[text=attackred,font=\small\bfseries]above:Loopjacking boundary}] {};
\end{tikzpicture}

%% file: figures/variants.tex
\begin{tikzpicture}[
  >=Latex,
  every node/.style={font=\small},
  stagebox/.style={draw, rounded corners=2pt, align=center, minimum height=9mm, text width=32mm, fill=softgray},
  alert/.style={stagebox, draw=attackred, very thick, fill=red!4},
  secure/.style={stagebox, draw=safegreen, very thick, fill=green!5},
  flow/.style={->, thick},
  bad/.style={->, very thick, draw=attackred},
  safe/.style={->, very thick, draw=safegreen}
]
  \node[font=\bfseries] at (1.975,1.1) {Representation-based};
  \node[stagebox] (r0) at (0,0) {Complete request\\already contains \(\B\)};
  \node[stagebox] (r1) at (0,-1.6) {Human sees only \(\A\)\\and approves it};
  \node[alert] (r2) at (0,-3.2) {Product uses approval\\for \(\A\); releases \(\B\)};
  \node[secure] (r3) at (3.95,-1.6) {Compare approved\\\(\A\) with complete \(\B\);\\reject or reauthorize};
  \draw[flow] (r0) -- (r1);
  \draw[bad] (r1) -- (r2);
  \draw[safe] (r1.east) -- (r3.west);
  \node[below=1mm of r3, text=safegreen, font=\bfseries\small,
        text width=32mm, align=center] {Safe path};

  \draw[densely dashed, gray] (6.15,1.2) -- (6.15,-4.1);

  \node[font=\bfseries] at (10.325,1.1) {Post-approval state substitution};
  \node[stagebox] (s0) at (8.35,0) {Human reviews and\\approves exact \(\A\)};
  \node[stagebox] (s1) at (8.35,-1.6) {Pending state changes\\from \(\A\) to \(\B\)};
  \node[alert] (s2) at (8.35,-3.2) {Stale approval for \(\A\)\\releases \(\B\)};
  \node[secure] (s3) at (12.30,-1.6) {Deny the change or\\revalidate current\\effect};
  \draw[flow] (s0) -- (s1);
  \draw[bad] (s1) -- (s2);
  \draw[safe] (s1.east) -- (s3.west);
  \node[below=1mm of s3, text=safegreen, font=\bfseries\small,
        text width=32mm, align=center] {Safe path};
\end{tikzpicture}

%% file: figures/a2a-carrier.tex
\begin{tikzpicture}[
  >=Latex,
  every node/.style={font=\small},
  stagebox/.style={draw, rounded corners=2pt, align=center, minimum height=12mm,
    fill=softgray},
  a2abox/.style={stagebox, draw=approvalblue, very thick, fill=blue!4},
  implbox/.style={stagebox, draw=black!65},
  alert/.style={stagebox, draw=attackred, very thick, fill=red!4},
  secure/.style={stagebox, draw=safegreen, very thick, fill=green!5},
  flow/.style={->, thick},
  bad/.style={->, very thick, draw=attackred},
  safe/.style={->, very thick, draw=safegreen}
]
  \node[a2abox, text width=29mm] (task) at (0,0)
    {\textbf{A2A}\\\textbf{coordination}\\Task \(T\) carries proposal \(\A\)};
  \node[implbox, text width=33mm, right=6mm of task] (review)
    {\textbf{Implementation / issuer}\\Human sees exact \(\A\);\\records \(D_A\)};
  \node[a2abox, text width=39mm, right=6mm of review] (update)
    {\textbf{A2A coordination}\\Authorization pending;\\message on \(T\) carries \(\B\);\\credential \(\rightarrow\) resume};
  \node[implbox, text width=31mm, right=6mm of update] (select)
    {\textbf{Implementation}\\Derives current\\operation \(\B\)};

  \node[alert, text width=43mm] (unsafe) at (9.8,-2.35)
    {\textbf{Unsafe composition}\\Consume \(D_A\) for \(\B\);\\release \(\B\)};
  \node[secure, text width=43mm] (safecheck) at (14.8,-2.35)
    {\textbf{Safe composition}\\Compare \(\B\) with \(D_A\);\\reject or reauthorize};

  \draw[flow] (task) -- (review);
  \draw[flow] (review) -- (update);
  \draw[flow] (update) -- (select);
  \draw[bad] (select.south west) to[out=-105,in=75] (unsafe.north);
  \draw[safe] (select.south east) to[out=-75,in=105] (safecheck.north);
\end{tikzpicture}

%% file: sections/cases.tex
\section{Loopjacking Cases}
\label{sec:cases}

We selected products to test the definition against different ownership and state
boundaries. The set contains two post-approval paths, one representation path, and
one implementation expected to preserve binding. Selection was purposive: the
cases were chosen because they expose a product-owned approval and a measurable
authorization-to-effect path, not to estimate how often the weakness occurs.

Across the experiments, actions were harmless but materially distinct. For the
state-substitution cases, \A\ was a mock transfer of 20 units to an approved vendor
and \B\ was a mock transfer of 2,000 units to an attacker-designated sink. The
tools appended their exact received arguments to a ledger. For the shell case,
\B\ created a temporary marker through a command that was not visible in \A.
These sinks establish the operation dispatched by the product without contacting
production services or moving real assets.

\begin{table}[!htbp]
\centering
\small
\caption{Admission of the three positive product paths. ``Direct \B\ denied''
means the mismatch was necessary for the attacker to obtain the tested effect.
The LangGraph result is conditional on the stated authorization policy.}
\label{tab:admission}
\begin{tabularx}{\textwidth}{@{}P{25mm}P{38mm}YY@{}}
\toprule
Product path & Approval and mismatch & Attacker boundary & Product consumption and controls \\
\midrule
Agno AgentOS regular Agent & Post-approval substitution. View: \code{transfer(20, vendor)}.
After admin approval, maker supplies \code{transfer(2000, sink)} &
Maker has run/continue authority, but no approval token or direct path to \B &
AgentOS installs and dispatches \B. Wrong actors and direct \B\ are denied; unchanged \A\ and the exact-action control pass \\
\addlinespace
LangGraph Agent Server, in-memory composition & Post-approval substitution. View: exact 20-unit
transfer. Maker replaces the pending same-ID call with a 2,000-unit transfer &
Maker may update shared state, but cannot resume or execute the transfer &
Approver resume dispatches \B. Direct \B, maker resume, and outsider fail; unchanged \A\ and deny-update policy pass \\
\addlinespace
OpenClaw shell-wrapper path & Representation mismatch. View: inline payload \code{\$0 "\$1"};
the pre-existing positional vector already encodes \B &
Requester cannot run \B\ without consuming an allow-once decision &
Gateway/node-host releases the full vector. Direct \B\ is denied, unchanged \A\ passes, and the fixed release rejects the mismatch \\
\bottomrule
\end{tabularx}
\end{table}
\FloatBarrier

\subsection{Agno AgentOS: approval followed by changed continuation}
\label{sec:case-agno}

\paragraph{Boundary.}
The tested deployment used a regular Agno Agent behind AgentOS's HTTP API. JSON
Web Tokens represented three relevant roles: a maker with \code{agents:run}, a
separate approval administrator with \code{agent\_os:admin}, and actors without
the relevant task authority. The maker could start and continue its own run but
could not resolve an approval. The administrator could resolve the product's
approval record but did not supply the continuation. This split matters: the
attack does not assume that the maker possesses the administrator token.

\paragraph{Trace.}
The maker first started \A, causing AgentOS to persist a required approval for the
exact tool name and arguments. The approval record associated the run with the
maker and showed only the 20-unit transfer to the approved vendor. Pending-state
and direct-\B\ controls produced no \B\ ledger event. The approval administrator
then resolved \A; the stored record named the administrator as resolver and still
contained only \A. After the pending approval cleared, the maker continued the same
run and supplied a tool execution with the original call identity but arguments
for \B\ and a confirmed state. AgentOS accepted the continuation and invoked the
registered tool with \B. No approval for \B\ appeared in the approval list.

The tested source path explains the result. The continuation route first checks
whether a required approval remains pending. Once that run-level state clears, it
deserializes caller-supplied tool executions, installs the supplied object into the
run, and dispatches a confirmed tool execution. In the main tested path, call
identity is matched, but the current tool arguments are not compared with the
descriptor in the approved record. The missing binding is therefore between the
approved pair \((\mathrm{tool},\mathrm{args})\) and the pair used at dispatch.

On the exact regular-Agent configuration, released Agno 2.5.6, 2.9.0, 3.0.1,
3.0.2, 3.0.3, 3.0.6, and 3.0.9 produced this trace. The adjacent 2.5.5 control
allowed direct \B, so it is not a qualifying Loopjacking result; 2.5.6 is the
established lower boundary at which the approval gate both blocked direct \B\ and
could be reused for substituted \B.\claim{C102} Package release metadata confirms
the evidence-freeze endpoint used here \cite{AgnoPyPI2026}. We do not infer that
every intermediate patch is affected, and no fixed Agno release is established.

\paragraph{Controls.}
The matrix denied a maker's attempt to approve, rejected wrong-user, outsider,
pending-state, and wrong-session operations, and allowed unchanged approved \A.
A research control that stored the approved canonical descriptor and compared it
at use time rejected \B\ while preserving \A. That control demonstrates the
repair invariant but is not an Agno product fix.

\subsection{LangGraph Agent Server: conditional shared-thread substitution}
\label{sec:case-langgraph}

\paragraph{Boundary.}
The LangGraph experiment used the released Agent Server with its in-memory runtime,
LangChain 1.3.18, LangGraph 1.2.11, and a supported custom authorization policy.
The policy gave a maker permission to update a shared pending thread, but not to
resume it or invoke the protected transfer. A distinct approver could inspect and
resume the thread and held the mock \code{wire:execute} permission, but could not
perform the maker's update. An outsider was denied. This is a plausible composition
of documented extension points, but it is not asserted to be a default deployment.

\paragraph{Trace.}
The maker created a thread whose interrupt displayed exact \A. The approver read
that interrupt. Before the approver acted, the maker used Agent Server's shipped
A2A message-command surface to submit \code{message.command.update}; the update
replaced the pending same-ID tool call with \B. The maker still could not resume the
thread or directly execute \B. The approver then submitted only the stale
\code{approve} decision derived from the earlier view. Agent Server resumed from
current thread state, and the tool received exact \B\ under the approver's identity
and permission. The attack ledger contained \B\ and not \A; a separate unchanged
control executed \A.

The product path crosses four components: Agent Server accepts and forwards the
message command; the message reducer replaces a same-ID pending message; HITL
processing reconstructs the current call when resuming; and execution inherits the
authenticated resume principal. No comparison binds the action displayed earlier
to the current same-ID action. The custom authorization policy creates the
separation of duties, but the replacement and resume behavior are supplied by the
released product composition.

The strict trace executed on 12 released Agent Server versions: 0.7.5, 0.7.103,
0.8.7, 0.9.1, 0.10.3, 0.11.4, 0.12.4, 0.12.6, 0.12.9, 0.13.2, 0.13.4, and 0.14.0.
Version 0.7.4 lacked the required command-update path. A hash-verified scan of 130
stable wheels found the required forwarding path in every release from 0.7.5
through 0.13.4; only the listed sampled versions were executed, and 0.14.0 is a
separate current-release execution point. Released LangGraph Agent Server therefore
executes \B\ in the tested in-memory composition when authorization permits a
non-approver to replace the pending shared-thread action. A supported policy that
denies that update prevents \B\ and preserves \A, so the result is conditional,
not universal.\claim{C103} The current release point at the cutoff is independently
identified by package metadata \cite{LangGraphAPIPyPI2026}.

\paragraph{Controls and scope.}
Direct \B, maker resume, and outsider access did not produce the protected effect.
The unchanged path produced exact \A. In the safe-policy control, the same Auth
extension rejected the maker's update, state remained \A, and the approver executed
\A. Production Postgres was not evaluated: the official image reached its
license-key requirement before serving the client. We therefore make no Postgres
claim and identify no vendor-fixed release.

\subsection{OpenClaw: incomplete shell representation}
\label{sec:case-openclaw}

OpenClaw exposes a gateway-mediated approval path for commands dispatched to a node
host. In the affected shell-wrapper path, the request carried a positional argument
vector whose inline payload was \code{\$0 "\$1"}; additional positional arguments
selected a command and destination that created the \B\ marker. The full vector was
prepared before approval. The product approval event, however, represented only
the inline payload \A. After an allow-once decision was recorded through the
product approval-resolution path, version 2026.2.23 released the complete vector
to the node host and created the marker. Direct \B\ without approval was denied,
and an unchanged approved command succeeded.

OpenClaw 2026.2.24 added consistency checking over the complete argument vector.
Under the same input, it rejected the incomplete raw-command representation before
node execution, produced no \B\ marker, continued to deny direct \B, and preserved
unchanged \A. Thus released 2026.2.23 consumed approval for incomplete \A\ and
executed complete \B, while 2026.2.24 rejected the mismatch.\claim{C104} The native
boundary agrees with the vendor advisory and patch
\cite{OpenClawGHSA6rcp2026,OpenClawFix0f0a2026}.

A later advisory describes substantively the same shell-wrapper mismatch but lists
a May boundary \cite{OpenClawGHSA2j8v2026}. Registry and commit-ancestry checks
found no stable 2026.5.16 artifact, and the checked May releases already contained
the February fix. We therefore use the original advisory and reproduced
2026.2.23/2026.2.24 boundary. Treating the later record as a duplicate is our
evidence-led reconciliation, not a formal vendor or GitHub merge decision.

The approval decision in these experiments was scripted only after the exact
product event had been asserted and recorded. The experiment establishes the
product's representation and release behavior. It does not establish how often a
person would understand or approve the same dialog.

%% file: sections/evaluation.tex
\section{Evaluation}
\label{sec:evaluation}

\subsection{Method}

For each product path we recorded the full request, the human-visible approval
representation, the decision, relevant task or call lineage, the operation at use
time, and the final ledger or marker effect. Trials used released packages pinned
in isolated environments. Networked products crossed loopback HTTP through their
documented server paths. Models were deterministic local fixtures because model
choice is not the claimed cause. Credentials were synthetic, and all external
effects were replaced by append-only ledgers or temporary files.

An attack result had to satisfy five controls in addition to observing \B: exact
\A\ was presented for approval; unchanged approved \A\ succeeded; direct \B\ under
the attacker identity failed; an actor outside the declared scope failed where the
product exposed that scope; and a fixed release or safe configuration blocked the
mismatch while preserving \A. We inspected the sink arguments, not merely an HTTP
status, a pending state, or a model response.

The retained evidence includes package manifests, dependency versions, raw
requests and responses, authorization decisions, approval records, final effects,
and file hashes. The Agno current-release matrix used Python 3.12.11 on macOS and
crossed its AgentOS HTTP path; earlier representative runs also used a clean Linux
container. LangGraph's 0.13.2 point was repeated on Linux; the release matrix and
the other current points were run on macOS. OpenClaw crossed its loopback gateway
and node-host path. OpenAI Agents used native \code{RunState} serialization and a
shipped deterministic test model.

\subsection{Results}

\begin{table}[!htbp]
\centering
\small
\caption{Attack and control outcomes. Counts are deterministic trial outcomes in
the cited frozen bundles, not population estimates. ``Safe'' names either a fixed
release, a supported safe policy, a research exact-action control, or a product
negative control as specified in the final column.}
\label{tab:outcomes}
\begin{tabularx}{\textwidth}{@{}P{27mm}P{28mm}P{22mm}P{22mm}Y@{}}
\toprule
Path & Tested points & Variant attack & Direct \B & Safe/fixed outcome \\
\midrule
Agno AgentOS & 2.5.6, 2.9.0, 3.0.1, 3.0.2, 3.0.3, 3.0.6, 3.0.9 &
Strict positive at each; 3.0.9: \AgnoAttackObserved/\AgnoAttackTotal &
3.0.9 denied \AgnoDirectBDenied/\AgnoDirectBTotal &
Research exact-action comparison rejected \B\ and preserved \A \\
\addlinespace
LangGraph Agent Server & 12 native positives from 0.7.5 through 0.14.0; 0.7.4 feature absent &
Strict positive at each admitted point &
Denied in current trace &
Supported deny-update Auth policy rejected mutation and executed \A \\
\addlinespace
OpenClaw & 2026.2.23 affected; 2026.2.24 fixed &
2026.2.23: \OpenClawAffectedObserved/\OpenClawAffectedTotal &
Denied \OpenClawAffectedDirectBDenied/\OpenClawAffectedTotal{} affected and
\OpenClawFixedDirectBDenied/\OpenClawFixedTotal{} fixed &
2026.2.24 rejected mismatch \OpenClawFixedObserved/\OpenClawFixedTotal; \A\ still executed \\
\addlinespace
OpenAI Agents SDK & 0.22.0 and 0.22.2 &
Mutation rejected \OpenAIEarlierMutationRejected/\OpenAIEarlierMutationTotal{} and
\OpenAICurrentMutationRejected/\OpenAICurrentMutationTotal &
Paused without effect &
Unchanged serialized \A\ executed \OpenAIEarlierHonestPass/\OpenAIEarlierHonestTotal{} and
\OpenAICurrentHonestPass/\OpenAICurrentHonestTotal \\
\bottomrule
\end{tabularx}
\end{table}
\FloatBarrier

\paragraph{Agno.}
Every listed strict-positive Agno release executed the substituted action in five
of five attack trials, executed unchanged \A\ in five of five controls, denied
three of three direct-\B\ attempts, passed the actor-boundary checks, and passed
the exact-action research control. The evidence-freeze current 3.0.9 bundle
completed all \AgnoCompletedTrials\ declared cells. Version 2.5.5 demonstrates
why the direct-path control is necessary: it also executed the continuation, but
because direct \B\ succeeded, approval hijacking was not needed. These results
establish the exact tested release points and the adjacent 2.5.5/2.5.6 boundary,
not an uninterrupted affected range.\claim{C102}

\paragraph{LangGraph.}
Each of the 12 admitted release compositions satisfied the strict trace: exact \A\
was reviewed, direct \B\ and maker resume were denied, the permitted update made
pending state exactly \B, approver resume executed exact \B, and an unchanged
control executed \A. The 0.7.4 negative lacked the required product feature. The
safe policy then falsified the universal claim by denying the maker's update while
leaving \A\ available for approval and execution. The source scan supports the
conditional continuity of the forwarding path through 0.13.4, but cannot promote
unexecuted patches to native results.\claim{C103}

\paragraph{OpenClaw.}
All \OpenClawAffectedObserved/\OpenClawAffectedTotal\ trials on 2026.2.23 captured the incomplete approval
view, the complete pre-existing vector, a product allow-once decision, and the \B\
effect. Each trial also denied direct \B\ and executed unchanged \A. All
\OpenClawFixedObserved/\OpenClawFixedTotal\ corresponding trials on 2026.2.24 rejected the mismatch
before node execution while retaining both controls. This paired release test
localizes the representation failure and validates the product fix.\claim{C104}

\paragraph{OpenAI Agents negative control.}
For each of versions 0.22.0 and 0.22.2, every approved exact-\A\ state was
serialized, restored, and resumed; all executed \A. In every attack trial, the
raw serialized copy of the pending invocation was changed to \B\
while retaining the call ID and original canonical approval/invocation records.
Restoration succeeded, but resume raised \code{ModelBehaviorError} and the ledger
remained empty. Direct unapproved \B\ paused. Ordinary per-call approval therefore
remained bound to the canonical invocation in these versions, demonstrating that
continuation and serialization are insufficient by themselves.\claim{C105} The
approval and resume model is documented by the SDK
\cite{OpenAIAgentsHITL2026}, and package metadata identifies the release endpoint
\cite{OpenAIAgentsPyPI2026}.

\subsection{Comparative interpretation}

The three positive paths fail at different stages. OpenClaw loses information
before the decision: the human-facing string is not the complete operation the node
will execute. Agno presents the correct action, but later treats the absence of a
pending run-level approval as sufficient for caller-supplied current arguments.
The LangGraph composition also presents the correct action, then permits a distinct
state writer to replace it before an approver resumes from current state. The
OpenAI control preserves a canonical invocation record across the same broad
pause/serialize/resume lifecycle and rejects the changed raw invocation.

This comparison supports two conclusions within the measured population. First,
resumability is a carrier for post-approval substitution but neither a necessary
condition for the umbrella nor a sufficient cause of failure. Second, the secure
object must be the complete current operation, not only a call identifier, a run's
approval status, or a rendered substring. Across the tested traces, complete
canonical rendering plus use-time comparison, or preventing unauthorized pending
state mutation, blocks \B\ without breaking \A.\claim{C107}

%% file: sections/related-work.tex
\section{Distinguishing Loopjacking}
\label{sec:related}

The underlying security rule is established: the action a person approves should
be the action that executes. Consent Integrity defines trusted
rendering and bind-to-execution for black-box agent approval
\cite{Weng2026ConsentIntegrity}. SUDP binds a fresh authorization grant to a
canonical operation and treats substitution and replay as protocol threats
\cite{YuEtAl2026SUDP}. Microsoft's action-bound approval design similarly specifies
durable action digests, authenticated approver chains, expiry, consumption, and
execution-time revalidation \cite{Microsoft2026ActionBoundApproval}. Loopjacking's
contribution is therefore an operational definition, a two-variant comparison, and
evidence from named released product paths; it is not the invention of approval
integrity, exact-action grants, or revalidation.\claim{C101}

\begin{table}[H]
\centering
\footnotesize
\caption{Nearest-neighbor distinctions. The decisive Loopjacking condition is
product-owned consumption of a genuine human decision for \A\ to authorize or
release materially different \B.}
\label{tab:neighbors}
\begin{tabularx}{\textwidth}{@{}P{31mm}YYY@{}}
\toprule
Neighbor & Core mechanism & Relationship & Missing or distinguishing condition \\
\midrule
Lies in the Loop \cite{OWASP2026LiesInTheLoop} &
Approval-dialog content or presentation is manipulated &
Direct overlap with representation-based Loopjacking &
Loopjacking also covers correct review followed by product-owned state substitution \\
\addlinespace
Consent Integrity \cite{Weng2026ConsentIntegrity} &
Trusted rendering and bind-to-execution &
Directly states the security property and covers swap threats &
This paper contributes product comparison and variant classification, not the property \\
\addlinespace
Agent Session Smuggling \cite{Unit422026SessionSmuggling} &
A remote agent injects covert instructions into an active cross-agent session &
Can deliver \B\ into a stateful workflow &
Needs no action-specific human decision for \A\ or consumption of that decision for \B \\
\addlinespace
Authorization Continuity \cite{ZhangZhang2026AuthorizationContinuity} &
Determines whether a live grant survives changes to agent state, delegation, or task phase &
Closest general stale-grant model &
Loopjacking requires the concrete human-approval-to-effect trace and reachable mismatch actor \\
\addlinespace
AID-Guard \cite{TongEtAl2026AIDGuard} &
Revalidates requests and provider state at commit, retry, and recovery &
Direct overlap with post-admission mutation and repair &
Not itself a product approval-hijacking instance \\
\addlinespace
Memory poisoning \cite{DashEtAl2026MemoryPoisoning} &
Untrusted data persists and steers later behavior &
Can supply or influence \B &
Requires neither approval for \A\ nor reuse of its authority \\
\bottomrule
\end{tabularx}
\end{table}
\FloatBarrier

\subsection{Representation and approval deception}

OWASP's Lies in the Loop names attacks that manipulate approval-dialog content so
a harmful operation appears benign \cite{OWASP2026LiesInTheLoop}. That work is a
direct neighbor, and some of its cases satisfy our representation variant when the
product's approval view omits or misstates a material part of the operation. The
scope is not identical: Loopjacking requires the resulting \A-to-\B\ authorization
trace and also includes a correct approval view followed by state substitution.
Ordinary social engineering, where a person knowingly approves fully visible \B,
is outside our definition.

OpenClaw's advisory is one concrete representation example. Other public records
show different approval-to-effect mismatches. Cursor's MCPoison advisory describes
a benign MCP configuration being approved and later changed by a repository writer
without renewed approval \cite{Cursor2025MCPoison}. We use this as a
source-grounded analogue, not as a locally reproduced case or evidence of
prevalence.

\subsection{Changing state and authorization continuity}

Recent agent-security work directly addresses authorization under changing state.
Authorization Continuity asks whether a grant remains valid when an agent's state,
tools, delegation, enforcement, or task phase changes
\cite{ZhangZhang2026AuthorizationContinuity}. Stateful Governance requires effects
to serialize against current policy and resource state
\cite{PengWu2026StatefulGovernance}. AID-Guard extends revalidation across mutation,
retry, ambiguous outcomes, and recovery \cite{TongEtAl2026AIDGuard}. Earlier and
non-agent state-divergence examples likewise show that checking one representation
and using another can grant unintended authority
\cite{TrailOfBits2026StateDivergence}. These works preclude a novelty claim over
stale authorization or commit-time checking. Our narrower question is how a human
approval becomes attached to a different concrete effect in released agent
products, and which stage produces the mismatch.

Action-governance systems supply complementary prevention. AgentBound binds policy
and governance receipts to specific actions
\cite{KaulEtAl2026AgentBound}, while deterministic pre-action authorization checks
each structured tool call before execution
\cite{Uchibeke2026PreActionAuthorization}. Such a check prevents the tested attacks
when it reconstructs the complete current operation, evaluates the right policy,
and fails closed on any unapproved change.

\subsection{Session, prompt, and memory attacks}

Agent Session Smuggling injects covert multi-turn instructions through an active
cross-agent session \cite{Unit422026SessionSmuggling}. It can change what an agent
attempts, but it does not require a human approval or reuse an earlier decision.
Session smuggling may therefore provide \B\ to a Loopjacking-prone workflow; the
two classifications compose rather than subsume one another. Memory poisoning is
similar in this respect: durable untrusted data can influence a later operation,
but the attack becomes Loopjacking only if product approval for \A\ is then used
for that changed operation \cite{DashEtAl2026MemoryPoisoning}.

Prompt injection and unsafe allowlists can also bypass confirmation without
hijacking a particular human decision. Published agent command-execution work
includes paths classified as safe or pre-approved even when injected arguments
make them harmful \cite{TrailOfBits2025PromptRCE}. Google ADK Python's public
confirmation-continuation issue similarly accepts forged confirmation state rather
than consuming a genuine human decision for \A\ \cite{GoogleADK2026Confirmation}.
Both are security-relevant and motivate exact current-operation checks, but neither
is counted as a Loopjacking result under our definition.

%% file: sections/defenses.tex
\section{Detection and Prevention}
\label{sec:defenses}

The approval-binding invariant illustrated in \cref{fig:approval-model} gives the
defense: preserve the operation the human approved, reconstruct the operation that
will execute, and compare their complete canonical descriptors at the last
authorization point. The comparison must occur after all attacker-influenced
parsing, templating, state reduction, continuation, default insertion, wrapper
expansion, and argument resolution.

\subsection{Canonical approval records}

An approval record should bind at least:

\begin{itemize}[leftmargin=*,itemsep=2pt]
  \item the canonical action or tool identity and every material argument;
  \item the target resource, destination, and relevant side-effect class;
  \item the initiating principal and the task, thread, or session scope;
  \item the approving principal and the policy or issuer that interpreted the
        request;
  \item a nonce, creation time, expiry, and consumption status; and
  \item a digest over the same complete descriptor shown to the human.
\end{itemize}

The display must be generated from this descriptor, not from a lossy convenience
string while execution uses a richer object. Shell wrappers deserve particular
care: inline payloads, positional arguments, environment changes, working
directory, interpreter flags, and redirections can all change the effect. A
collapsed view may remain useful, but approval must expose any material hidden
fields or refuse to authorize them.

\subsection{Use-time decision}

Immediately before releasing the effect, the enforcement point should reconstruct
the complete resolved operation and compare it with the descriptor or digest bound
to the approval. A compact decision procedure is:

\begin{enumerate}[leftmargin=*,itemsep=2pt]
  \item reject if the decision is missing, expired, revoked, consumed, or outside
        its principal and task scope;
  \item reject or request new approval if the current operation differs materially
        from the approved descriptor;
  \item re-evaluate current policy over the resolved effect, including resource
        and destination constraints;
  \item atomically consume the decision with release of the effect, or record an
        idempotent commit token that prevents replay; and
  \item append the approved descriptor, current descriptor, decision, and effect
        to an audit record.
\end{enumerate}

Binding only a call identifier is insufficient when a reducer can replace the
arguments associated with that identifier. Binding only a run-level ``approved''
flag is insufficient when a continuation can install a different tool execution.
Checking only at presentation time is insufficient when state can change before
dispatch.

\subsection{Control placement}

Two control strategies blocked the studied paths.

\paragraph{Bind and revalidate.}
OpenClaw 2026.2.24 reconstructs and checks the complete shell argument vector,
rejecting the representation mismatch before node execution while still executing
unchanged \A. OpenAI Agents' ordinary per-call approval retains a canonical
invocation fingerprint across serialization and rejects a same-call-ID raw
invocation change. A research exact-action wrapper around the Agno trace likewise
rejected \B\ and preserved \A. This is the more general strategy because it
survives authorized or benign state changes by demanding a fresh decision whenever
the material effect changes.

\paragraph{Prevent unauthorized mutation.}
The LangGraph safe policy denied a non-approver's pending-thread update. The action
remained exact \A\ and the approver executed \A. This is effective when business
roles can forbid such updates. It is composition-specific, however, and it does
not replace exact-action binding when legitimate updates, retries, or migrations
can change operation state.

For the tested traces, complete canonical approval rendering plus exact
current-effect checking, or a policy that prevents the unauthorized pending-state
change, blocked \B\ while preserving legitimate \A.\claim{C107} This evidence is
bounded: OpenClaw supplies a released product fix; the LangGraph policy is a
supported safe configuration; the Agno wrapper is a research control rather than a
released fix. The design aligns with prior consent-integrity, operation-binding,
and action-bound approval proposals
\cite{Weng2026ConsentIntegrity,YuEtAl2026SUDP,Microsoft2026ActionBoundApproval}.

\subsection{Testing guidance}

A regression test should preserve more than the final status code. It should
capture the approval view and complete request, mutate one material field through
each supported post-review surface, and assert the exact operation at the sink.
The minimum suite is: unchanged \A\ succeeds; denial has no effect; direct \B\ is
unavailable to the attacker; \A-to-\B\ representation or state substitution is
rejected or reauthorized; wrong scope fails; and consumed or expired approval
cannot be replayed. Testing both variants avoids fixing only the visible symptom.

%% file: sections/limitations.tex
\section{Limitations, Ethics, and Reproducibility}
\label{sec:limitations}

\subsection{Study limits}

The four product paths form a purposive set, not a random or representative sample.
They establish mechanically distinct examples and a negative control, but cannot
measure ecosystem prevalence or show that most agent frameworks are affected. The
results are configuration- and version-specific. No universal CWE, CVSS score,
severity, or impact estimate follows from the umbrella definition.

Agno was executed at seven qualifying release points plus the ineligible 2.5.5
boundary control. Intermediate patches were not all executed or source-scanned,
so we do not claim one uninterrupted affected range. No fixed Agno release is
known at the cutoff, and its exact-action control in this study is
research-authored.\claim{C102}

The LangGraph result depends on the in-memory runtime, LangChain 1.3.18, LangGraph
1.2.11, and an Auth policy that allows a non-approver to update a shared pending
thread. A supported deny-update policy is safe, so the result does not characterize
all deployments. The wheel scan verifies the required forwarding path through
0.13.4, while 0.14.0 is a separate native point. Production Postgres behavior was
not tested because the official setup required a deployment or enterprise license
key, and no vendor fix is established.\claim{C103}

OpenClaw is the only studied positive path with a native affected/fixed release
pair. Its approval role was automated after checking the exact product approval
event. This isolates binding behavior but does not measure user comprehension,
deception rate, or interface quality. The later advisory's duplicate status is an
evidence-led reconciliation rather than a formal vendor determination.\claim{C104}

The OpenAI result is limited to ordinary function-tool per-call approval in 0.22.0
and 0.22.2. It does not cover intentional sticky \code{always\_approve} policy or an
attacker able to forge every trusted canonical field in serialized state. It is a
negative control, not a claim that all application-authored approval around the SDK
is safe.\claim{C105}

All experiments were operated by one researcher. Selected points were repeated in
clean macOS environments and, for representative Agno and LangGraph versions, in
Linux containers, but there is no independent reproduction. Our literature and
public-source review is current through September 10, 2026; we make no claim of
exhaustive coverage or priority.

\subsection{Ethics and disclosure}

Testing was confined to released software in local, isolated environments. All
network traffic used loopback transports. Tokens and identities were synthetic;
secrets were omitted from retained traffic. Consequential tools wrote only to mock
ledgers or temporary markers. We did not contact production systems, move funds,
access third-party data, or exercise destructive actions.

The study was conducted under coordinated-disclosure constraints. This manuscript
uses public advisories, public specification changes, and the product evidence
described here; it excludes private coordination messages, personal contact data,
unverified chronology, and unrelated vulnerability tracks. Product findings are
kept separate: the A2A specification history, Agno configuration, conditional
LangGraph composition, OpenClaw advisory, and OpenAI negative control do not share
one asserted root cause.

\subsection{Reproducibility}

The accompanying evidence archive \cite{ArunKumar2026LoopjackingEvidence}
records, for each admitted point, the released
package version, dependency pins, platform, test command, exact actors and scopes,
requests, approval representation, canonical \A\ and \B, decision, raw transport
or serialized state, final ledger or marker, and checksums. Running
\texttt{python3 verify\_archive.py} from the repository root recomputes the retained
bundle hashes and the declared result oracles.
Generated manuscript values are read from frozen result objects rather than typed
into the results table.

Reproduction does not require external side effects. The Agno and LangGraph tests
start local servers and use deterministic models; the OpenClaw test connects a
local gateway and node host; the OpenAI control serializes and restores native run
state. A reproducer should verify both the attack and its paired controls and
should treat a missing licensed Postgres environment as outside the LangGraph
claim, not as a negative result.

%% file: sections/conclusion.tex
\section{Conclusion}
\label{sec:conclusion}

A human approval is not a security boundary unless it remains attached to the
complete operation that inherits its authority. Loopjacking names the product-owned
failure in which approval understood as \A\ authorizes or releases materially
different \B. The mismatch may exist before approval in an incomplete
representation or arise afterward through mutable workflow state.\claim{C101}

The evaluated products show both variants and their boundary conditions. The
tested Agno configuration and conditional LangGraph composition demonstrate
post-approval substitution; OpenClaw demonstrates pre-approval representation
mismatch; and OpenAI Agents provides a counterexample in which resumable serialized
state preserves action binding. These are exact product and configuration results,
not a prevalence estimate or a claim that approval integrity is new.

The practical rule is simple to state and easy to omit: render the complete
canonical operation, bind the decision to that descriptor and its scope, reconstruct
the actual current effect at use time, and reject or reauthorize any material
change. Preventing unauthorized pending-state mutation adds useful defense in
depth, but exact use-time binding remains the portable invariant. Those controls
blocked the studied attacks without breaking legitimate \A.\claim{C107}

%% file: sections/ai-use.tex
\section*{Generative AI Usage}

To help manage and cross-check the repository-scale evidence corpus, OpenAI Codex
assisted with repository inspection, orchestration and checking of experiment
scripts, evidence normalization, citation-metadata collection, drafting and
revision of manuscript text and \LaTeX{} figures, and build and visual verification.
The author determined the research questions, scope, methods, evidentiary
boundaries, claims, and conclusions; reviewed the cited sources, experimental
records, controls, citations, and final manuscript; and made all substantive
decisions. Codex outputs were not treated as independent evidence or validation.
The author takes full responsibility for the paper.